\documentclass[10pt]{article}
\usepackage[cp866]{inputenc}
\begin{document}

\bigskip

\centerline {{\Large\bf Skew-symmetric
forms: On integrability }}
\centerline {{\Large\bf of equations of mathematical physics.}}
\centerline {\it L.~I. Petrova}

\renewcommand{\abstractname}{Abstract}
\begin{abstract}

The study of integrability of the mathematical physics equations
showed that the differential equations describing real processes
are not integrable without additional conditions. This follows
from the functional relation that is derived from these equations.
Such a relation connects the differential of state functional and
the skew-symmetric form. This relation proves to be nonidentical,
and this fact points to the nonintegrability of the equations.  In
this case a solution to the equations is a functional, which
depends on the commutator of skew-symmetric form that appears to
be unclosed. However, under realization of the conditions of
degenerate transformations, from the nonidentical relation it
follows the identical one on some structure. This points out to
the local integrability and realization of a generalized solution.

In doing so, in addition to the exterior forms, the skew-symmetric 
forms, which, in contrast to exterior forms, are defined on nonintegrable 
manifolds (such as tangent manifolds of differential equations, Lagrangian 
manifolds and so on), were used. 

In the present paper, the partial differential equations, which describe
any processes, the systems of differential equations of mechanics and physics
of continuous medium and field theory equations are analyzed.

\end{abstract}

\section{Analysis of partial differential equations that describe real 
processes}
Let us take the simplest case: the  first-order partial differential equation 
$$ F(x^i,\,u,\,p_i)=0,\quad p_i\,=\,\partial u/\partial x^i \eqno(1)$$
We consider the functional relation
$$ du\,=\,\theta\eqno(2)$$
where $\theta\,=\,p_i\,dx^i$ is a skew-symmetric differential form of the first
degree (the summation over repeated indices is implied).

In the general case, when differential equation (1) describes any physical
processes, the functional relation  (2) is nonidentical one. If to take the
differential of this relation, we will have $ddu=0$ in the left-hand side,
whereas in the right-hand side $d\theta$ is not equal to zero.
The differential $d\theta$ is equal to $K_{ij}dx^idx^j$, 
where $K_{ij}=\partial p_j/\partial x^i-\partial p_i/\partial x^j$
is components of differential form commutator constructed of the mixed 
derivatives. From equation (1) it does not follow (explicitly) that the
derivatives $p_i\,=\,\partial u/\partial x^i $, which obey to the equation
(and given boundary or initial conditions) are consistent, and that mixed
derivatives are commutative. Components of commutator $K_{ij}$ is nonzero. 
Therefore the differential form commutator and the differential of 
form $\theta$ are nonzero.  Thus, $d\theta$ is not equal to zero. 

The nonidentity of functional relation (2) means that the equation (1)
is nonintegrable: the derivatives  $p_i$ of  equation  do not make
up a differential. The solution $u$ of the equation (1), obtained from
such derivatives, is not a function of only variables $x^i$. This
solution  will depend on the commutator $K_{ij}dx^idx^j$, that is, it is a
functional.

To obtain a solution that is a function (i.e., the derivatives of
this solution make up a differential), it is necessary to add the
closure condition for the form $\theta\,=\,p_idx^i$ and for
the relevant dual form (in the present case the functional $F$
plays a role of a form dual to $\theta $) [1]:
$$\cases {dF(x^i,\,u,\,p_i)\,=\,0\cr
d(p_i\,dx^i)\,=\,0\cr}\eqno(3)$$

If we expand the differentials, we get a set of homogeneous equations
with respect to $dx^i$ and $dp_i$ (in the $2n$-dimensional tangent
space):
$$\cases {\displaystyle \left ({{\partial F}\over {\partial x^i}}\,+\,
{{\partial F}\over {\partial u}}\,p_i\right )\,dx^i\,+\,
{{\partial F}\over {\partial p_i}}\,dp_i \,=\,0\cr
dp_i\,dx^i\,-\,dx^i\,dp_i\,=\,0\cr} \eqno(4)$$

It is well known that {\it vanishing the determinant} composed of coefficients
at $dx^i$, $dp_i$ is a solvability condition of the system of homogeneous
differential equations. This leads to relations:
$$
{{dx^i}\over {\partial F/\partial p_i}}\,=\,{{-dp_i}\over {\partial F/\partial x^i+p_i\partial F/\partial u}} \eqno (5)
$$

Relations (5) specify the integrating direction, namely, a pseudostructure,
on which the form $\theta \,=\,p_i\,dx^i$ turns out to be closed one,
i.e. it becomes a differential, and from relation (2) the identical relation
is produced.  One the pseudostructure, which is  defined by relation(5),
the derivatives of differential equation (1) constitute
a differential $\delta u\,=\,p_idx^i\,=\,du$ (on the pseudostructure), and
this means that the solution to equation (1) becomes a function.

Solutions, namely, functions on the pseudostructures
formed by the integrating directions, are the so-called generalized
solutions.

It is evident that this solution is obtained only under degenerate
transformation, that is, when the determinant vanishes.

{\footnotesize [It is evident that the degenerate transformation is a transition
from tangent space to cotangent space (the Legendre transformations).
The coordinates in relations (5) are not identical to the independent
coordinates of the initial space on which equation (1) is defined.]}

\bigskip
The first-order partial differential equation has been analyzed,
and the functional relation with the form of the first degree
has been considered.

Similar functional properties have all differential
equations describing real processes. And, if the order of the
differential equation is $k$, the functional relation with the
$k$-degree form corresponds to this equation.

Here the following  should be emphasized. Under degenerate
transformation from an initial nonidentical functional relation the 
integrable identical relation is obtained. As a
result of integrating, one obtains the relation that contains
skew-symmetric forms of degree less by one and which in turn
proves to be a nonidentical (without additional conditions). By
integrating the functional relations sequentially obtained (it is
possible only under realization of the degenerate
transformations), from the initial functional relation of degree
$k$ one can obtain $(k+1)$ functional relations each involving
exterior forms of one of degrees: $k, \,k-1, \,...0$. In
particular for the first-order partial differential equation it is
also necessary to analyze the functional relation of zero degree.

\bigskip
Thus one can see that  the nonintegrability of differential equations describing real processes
is due to the nonconjugacy (noncommutativity) of the derivatives with respect
to different variables: the commutator made up of relevant mixed derivatives is nonzero.
Without the realization of additional conditions, the solution will depend on
this commutator, that is, it will be a functional.
For ordinary differential equations the relevant 
commutator is generated due to the conjugacy of the derivatives
of the functions desired and those of initial data.

It should be emphasized once more that all differential equations describing
real physical processes allow  solutions of two types, namely,
generalized solutions that depend on variables only, and solutions
that are functionals since they depend on the commutator made up
by mixed derivatives.

{\footnotesize [The dependence of the solution on the commutator may lead to
instability of the solution. Equations that do not satisfy the integrability
conditions may have unstable solutions. Unstable
solutions appear in the case when the additional conditions are not
realized and no exact solutions (their derivatives make up a differential)
are formed. Thus, the solutions to the equations of the elliptic type may
be unstable.

Investigation of nonidentical functional relations lies at the basis of the
qualitative theory of differential equations. It is well known that the
qualitative theory of differential equations is based on the analysis of
unstable solutions and the integrability conditions. From the functional
relation it follows that the dependence of the solution on the commutator
leads to instability, and the closure conditions of skew-symmetric forms
constructed by derivatives are integrability conditions. That is,
the qualitative theory of differential equations that solves the problem of
unstable solutions and integrability bases on the properties of nonidentical
functional relation.]}

Thus one can see that the solutions to equations of mathematical physics,
on which no additional conditions are imposed, are functionals.
The solutions prove to be exact (the generalized solution) only under
realization of additional requirements, namely,  the conditions of degenerate
transformations. Mathematically, this corresponds to some functional expressions
that become equal to zero. Such functional expressions are Jacobians, determinants,
the Poisson brackets, residues, and others. These conditions define
integral surfaces. The characteristic manifolds, the envelopes of
characteristics, singular points, potentials of simple and double layers,
residues and others are the examples of such surfaces.

Since generalized solutions are possible only under realization of
the conditions of degenerate transforms, they are discrete
solutions (defined only on pseudostructures) and have
discontinuities in the direction normal to pseudostructures.

\bigskip
While studying the integrability of differential equation, which
was carried out by using the skew-symmetric differential forms, a
nontraditional mathematical apparatus made its appearance, namely,
the nonidentical relation and degenerate transformation. The
skew-symmetric differential forms, which, in contrast to interior
forms, are defined on nonintegrable manifolds (such, for example,
as the tangent manifolds of differential equations) and possess
the evolutionary properties, are provided with such mathematical
apparatus.

Below, the analysis of integrability of the equations of mechanics
and physics of continuous medium will be carried out by using the
exterior and evolutionary forms. In so doing, the mechanism of
realization the integrability of these equations will be disclosed
in more detail.

\section{Integrability of the equations of mechanics and physics of
continuous medium}
While studying the integrability of partial differential equations,
the conjugacy of derivatives with respect to different variables
was analyzed by using the nonidentical functional relation.
When describing physical
processes in continuous media (in material systems) one obtains not one
differential equation but a set of differential equations. And in
this case it is necessary to investigate the conjugacy of not only
derivatives with respect to different variables but also the conjugacy
(consistency) of the equations of this set. In this case, from
this set of equations one also obtains  a nonidentical relation that
enables one to investigate the integrability of equations and
the specific features of their solutions.

The equations of mechanics and physics of continuous media (of material systems)
is a set of equations that describe the conservation laws for energy, linear
momentum, angular momentum and mass. The Euler and Navier-Stokes equations are
examples of such a set of equations [1]. 

{\footnotesize [It should be noted that the conservation laws for
material systems are described with differential equations since
they are balance ones (they specify the balance between the
variations of physical quantities and external actions). In
contrast to this, the conservation laws for physical fields are
described by closed exterior forms since they are exact (point out
to the availability of conserved quantities).]}

Let us analyze the equations of energy and linear momentum.

In the accompanying reference system, which is tied to  the manifold
built by the trajectories of particles (elements  of material system), the
equation of energy is written in the form
$$
{{\partial \psi }\over {\partial \xi ^1}}\,=\,A_1 \eqno(6)
$$

Here $\xi ^1$ are the coordinates along the trajectory, $\psi $ is
the functional of the state that specifies material system, $A_1$ is
the quantity that
depends on specific features of the system and on external energy
actions onto the system.
{\footnotesize \{The action functional, entropy, wave function
can be regarded as examples of the functional $\psi $. Thus, the
equation for energy expressed in terms of the action functional $S$ has
a similar form:
$DS/Dt\,=\,L$, where $\psi \,=\,S$ and $A_1\,=\,L$ is the Lagrange function.
The equation for energy of an ideal gas can be presented in the form [1]:
$Ds/Dt\,=\,0$, where $s$ is entropy.\}}

In a similar manner, in the accompanying reference system the
equation for linear momentum appears to be reduced to the equation of
the form
$$
{{\partial \psi}\over {\partial \xi^{\nu }}}\,=\,A_{\nu },\quad \nu \,=\,2,\,...\eqno(7)
$$
where $\xi ^{\nu }$ are the coordinates in the direction normal to the
trajectory, $A_{\nu }$ are the quantities that depend on the specific
features of material system and on external force actions.

Eqs. (6) and (7) can be convoluted into the relation
$$
d\psi\,=\,A_{\mu }\,d\xi ^{\mu },\quad (\mu\,=\,1,\,\nu )\eqno(8)
$$
where $d\psi $ is the differential
expression $d\psi\,=\,(\partial \psi /\partial \xi ^{\mu })d\xi ^{\mu }$.

Relation (8) can be written as
$$
d\psi \,=\,\omega \eqno(9)
$$
here $\omega \,=\,A_{\mu }\,d\xi ^{\mu }$ is the skew-symmetrical differential
form of the first degree.

Relation (9) has been obtained from the equation of the balance conservation
laws for energy and linear momentum. In this relation the form $\omega $
is that of the first degree. If the equations of the balance conservation
laws for angular momentum be added to the equations for energy and linear
momentum, this form  will be a form of the
second degree. And, in combination with the equation of the balance
conservation law for mass, this form will be a form of degree 3.
In general case the evolutionary relation can be written as
$$
d\psi \,=\,\omega^p \eqno(10)
$$
where the form degree  $p$ takes the values $p\,=\,0,1,2,3$.
{\footnotesize (The relation for $p\,=\,0$ is an analog to
that in the differential forms, and it has been obtained from the
interaction of energy and time.)}

The relation obtained from the equation of the balance
conservation laws possess the properties that enable one to
investigate the integrability of the original set of equations.

This relation is, firstly, an evolutionary one since the original 
equations are evolutionary.

Secondly, it, as well as functional relation (2), turns out to be nonidentical.

To justify this we shall analyze the relation (9).

The evolutionary relation  $d\psi \,=\,\omega $ is a nonidentical relation as
it involves the unclosed differential form $\omega\,=\,A_{\mu }d\xi ^{\mu }$.
The commutator of the form $\omega $  is nonzero.
The components of the commutator of such a form can be written as follows:
$$
K_{\alpha \beta }\,=\,\left ({{\partial A_{\beta }}\over {\partial \xi ^{\alpha }}}\,-\,
{{\partial A_{\alpha }}\over {\partial \xi ^{\beta }}}\right )
$$
(here the term  connected with the manifold metric form
has not yet been taken into account).

The coefficients $A_{\mu }$ of the form $\omega $ have been obtained either
from the equation of the balance conservation law for energy or from that for
linear momentum. This means that in the first case the coefficients depend
on the energetic action and in the second case they depend on the force action.
In actual processes the energetic and force actions have different nature
and appear to be inconsistent. The commutator of the form $\omega $ constructed of
the derivatives of such coefficients is nonzero.
This means that the differential of the form $\omega $
is nonzero as well. Thus, the form $\omega$ proves to be unclosed and cannot be 
a differential.

The nonidentity of the evolutionary relation, as well as the nonidentity
of the functional
relation (2), means that the initial equations of conservation laws
are not conjugated, and hence, they are not integrable. The solutions to
these equations can be functional or generalized solutions. In this case, the
generalized solutions are obtained only under degenerated transformations.

The questions arise of how the conditions of degenerate transformation
are realized and how the degenerate transformation proceeds.

The evolutionary relation has a peculiarity that enables one to answer these
questions.

Since this relation is evolutionary and nonidentical, it turns out to be
a selfvarying one (it is an
evolutionary relation and it contains two objects one of which appears to be
unmeasurable and cannot be compared with another one, and therefore the
process of mutual variation cannot terminate).
The selfvarying evolutionary relation leads to  realization of the
conditions of  degenerate transformation.
Under degenerate transformation,
from nonidentical relation the relation that is identical
on pseudostructure is obtained.

If the conditions of degenerate transformation are realized, from
the unclosed evolutionary form with nonvanishing differential
$d\omega^p\ne 0 $,
one can obtain a differential form closed
on pseudostructure. The differential of this form equals zero. That is,
it is realized the transition

$d\omega^p\ne 0 \to $ (degenerate transformation) $\to d_\pi \omega^p=0$,
$d_\pi{}^*\omega^p=0$

The condition  $d_\pi{}^*\omega^p=0$ is an equation of a certain
pseudostructure $\pi$ on which  the differential of evolutionary
form vanishes: $d_\pi \omega^p=0$. That is, the closed
(inexact) exterior form $\omega_\pi^p$ is obtained on pseudostructure.
On the pseudostructure, from evolutionary relation $d\psi \,=\,\omega^p$
it is obtained an identical relation $d_\pi\psi=\omega_\pi^p$, since the
closed exterior form $\omega_\pi^p$ is a differential of some differential form
(this relation will be an identical one as the left and right sides 
of the relation contain differentials).
The identity of the relation obtained from the evolutionary relation
means that on pseudostructures the original equations for material systems
(the equations of conservation laws) become consistent and integrable.

Pseudostructures constitute the integral surfaces (such as characteristics,
singular points, potentials of simple and double layers and others)
on which the quantities of material system desired (such as the temperature,
pressure, density) become functions of only independent variables and do not
depend on the commutator (and on the path of integrating). This are generalized solutions.

One can see that the integral surfaces are obtained from the condition of
degenerate transformation of the evolutionary relation. As it was already mentioned,
the conditions of degenerate transformation are a vanishing of
such functional expressions as determinants, Jacobians, Poisson's
brackets, residues and others. They  are connected with the symmetries,
which can be due to the degrees of freedom (for example, the translational,
rotational and oscillatory freedom of the of material system).

The degenerate transformation is realized as the transition from the
noninertial frame of reference to the locally inertial system, i.e. the
transition from nonintegrable manifold (for example, tangent one)
to the integrable structures and surfaces (such as the
characteristics, potential surfaces, eikonal surfaces, singular
points).

\bigskip
Thus, one can see that the problem of integrability is based on 
the properties of nonidentical evolutionary relation.

Nonidentical evolutionary relation,
which is obtained from the
equations for material media (and from which identical relations with
closed forms describing physical fields are obtained), allows to understand
the specific features  of the field-theory equations as well.

\section{Analysis of the field-theory equations}
The field-theory equations are equations for physical fields.
As the physical fields are described by the closed exterior
(inexact) forms, it is obvious that solutions to the field-theory equations
must be closed exterior forms, i.e. to be differentials. And such
differentials, which are closed exterior forms, can be obtained
from the nonidentical evolutionary relation. This means that in field
theory the nonidentical evolutionary relation can play a role of the
field-theory equation.

One can see that there exists the correspondence between the field-theory
equations and the nonidentical evolutionary relation.

The nonidentical evolutionary relation is a relation for functionals such as
wave-function, action functional, entropy, and others.
The field-theory equations are those for such functionals.

Another correspondence relates to the peculiarity of field-theory equations.

The peculiarity of field-theory equations consists in the fact that all
these equations have the form of relations. They can be relations in
differential forms or in the forms of their tensor or differential
(i.e. expressed in terms of derivatives) analogs.

The Einstein equation is a relation in differential forms. This equation
relates the differential of the first degree form (Einstein's tensor)
and the differential form of second degree, namely, the energy-momentum
tensor. (It should be noted that Einstein's equation follows from
the differential form of third degree).

The Dirac equation relates  Dirac's \emph{ bra-} and \emph{ cket}- vectors,
which made up the differential form of zero degree.

The Maxwell equations have the form of tensor relations.

The Schr\H{o}dinger's equations have the form of relations
expressed in terms of derivatives and their analogs.

From the field-theory equations, as well as from the nonidentical evolutionary
relation, the identical relation, which contains the closed exterior form, is obtained.

The closed exterior forms or their
tensor or differential analogs, which are obtained from identical relations,
are solutions to the field-theory equations.

As one can see, from the field-theory equations it follows such identical
relation as

1) The Dirac relations made up of Dirac's \emph{ bra-} and
\emph{ cket}- vectors, which connect a closed exterior form of zero degree;

2) The Poincare invariant, which connects  closed exterior forms of
first degree;

3) The relations $d\theta^2=0$, $d^*\theta^2=0$ are those for closed exterior
forms of second degree obtained from Maxwell equations;

4) The Bianchi identity for gravitational field.

From the Einstein equation it is obtained the identical relation in the
case when the covariant derivative of the energy-momentum tensor
vanishes.

It is evident that all equations of existing field theories are in
essence the relations that connect skew-symmetric forms or their
analogs. It may be emphasized once more that the equations of
field theories have the form of relations for functionals such as
wave function (the relation corresponding to differential form of
zero degree), action functional (the relation corresponding to
differential form of first degree), the Pointing vector (the
relation corresponding to differential form of second degree). The
tensor functionals that correspond to Einstein's equation are
obtained from the relation connecting the differential forms of
third degree.

The nonidentical evolutionary relations derived from the equations for
material media, as it was already mentioned, unites the relations for all 
these functionals. This is, all equations of field theories are analogous 
to the nonidentical evolutionary relation.

From this it follows that the nonidentical evolutionary relation can
play a role of the equation of general field theory that discloses
common properties and peculiarities of existing equations of field
theory.

Can see that investigation of integrability of the field-theory equations 
also is based on the properties of nonidentical evolutionary relation.

1. Clark J.~F., Machesney ~M., The Dynamics of Real Gases. Butterworths,
London, 1964.

\end{document}